\documentclass[aip,jcp,amsmath,amssymb,floatfix,citeautoscript,reprint]{revtex4-1}

\usepackage{graphicx}
\usepackage{dcolumn}
\usepackage{gensymb}
\usepackage{bm}

\usepackage{braket}
\usepackage[usenames,dvipsnames]{color}

\usepackage{ragged2e}
\usepackage[labelsep=period,format=plain, justification=justified,font=footnotesize]{caption}
\usepackage{times}
\usepackage{txfonts}

\DeclareCaptionJustification{myjust}{\justifying}
\usepackage{xr}
\makeatletter
\newcommand*{\addFileDependency}[1]{
  \typeout{(#1)}
  \@addtofilelist{#1}
  \IfFileExists{#1}{}{\typeout{No file #1.}}
}
\makeatother

\newcommand*{\myexternaldocument}[1]{
    \externaldocument{#1}
    \addFileDependency{#1.tex}
    \addFileDependency{#1.aux}
}

\myexternaldocument{SI}

\begin{document}

\title{Developing machine-learned potentials to simultaneously capture the dynamics of excess protons and hydroxide ions in classical and path integral simulations}

\author{Austin O. Atsango}
\affiliation{Department of Chemistry, Stanford University, Stanford, California, 94305, USA}

\author{Tobias Morawietz}
\affiliation{Department of Chemistry, Stanford University, Stanford, California, 94305, USA}

\author{Ondrej Marsalek}
\affiliation{Faculty of Mathematics and Physics, Charles University, Prague, Czech Republic}

\author{Thomas E. Markland}
\email{tmarkland@stanford.edu}
\affiliation{Department of Chemistry, Stanford University, Stanford, California, 94305, USA}

\date{\today}

\begin{abstract}
The transport of excess protons and hydroxide ions in water underlies numerous important chemical and biological processes. Accurately simulating the associated transport mechanisms ideally requires utilizing {\it ab initio} molecular dynamics simulations to model the bond breaking and formation involved in proton transfer and path-integral simulations to model the nuclear quantum effects relevant to light hydrogen atoms. These requirements result in a prohibitive computational cost, especially at the time and length scales needed to converge proton transport properties. Here, we present machine-learned potentials (MLPs) that can model both excess protons and hydroxide ions at the generalized gradient approximation and hybrid density functional theory levels of accuracy and use them to perform multiple nanoseconds of both classical and path-integral proton defect simulations at a fraction of the cost of the corresponding {\it ab initio} simulations. We show that the MLPs are able to reproduce {\it ab initio} trends and converge properties such as the diffusion coefficients of both excess protons and hydroxide ions. We use our multi-nanosecond simulations, which allow us to monitor large numbers of proton transfer events, to analyze the role of hypercoordination in the transport mechanism of the hydroxide ion and provide further evidence for the asymmetry in diffusion between excess protons and hydroxide ions.
\end{abstract}

\maketitle
\normalsize

\section{Introduction}
Water's ability to autoionize and efficiently transport its ionization products---excess protons and hydroxide ions---through its hydrogen bond network is a fundamental characteristic that underlies multiple processes ranging from acid-base chemistry to the operation of proton exchange membrane fuel cells~\cite{Peighambardoust2010} and voltage-gated proton channels in biological cell membranes~\cite{DeCoursey2018}. Excess protons and hydroxide ions are known to diffuse via structural (Grotthuss-like) mechanisms~\cite{Marx2006}, which involve the making and breaking of chemical bonds through a series of proton transfer reactions between neighboring water molecules. These structural diffusion mechanisms allow both species to diffuse much faster than water itself and are intricately linked to the structure and dynamics of the hydrogen bonds that solvate proton defects in water~\cite{Hassanali2013}. Excess protons and hydroxide ions thus exhibit different diffusion rates in water due to their different solvation motifs, and nuclear magnetic resonance (NMR)~\cite{Halle1983,Halle1983-2} and conductivity~\cite{Halle1983,Weingrtner1990,Sluyters2010} measurements have shown that excess protons diffuse $\sim$1.8 times faster than hydroxide ions at room temperature. The need for a deeper understanding of the complex molecular structures and motions that lead to these diffusion mechanisms and the differences between the diffusion rates of excess protons and hydroxide ions has led to extensive theoretical studies~\cite{Tuckerman1995xa,Tuckerman1995xb,Agmon1995,Marx1999,Marx2000,Vuilleumier1999,Day2000,Tuckerman2002,Agmon2016}. However, due to the reactive nature of the defects, which necessitate a quantum mechanical treatment of the electrons to allow chemical bonds to be made and broken during the simulation, and their low mass, which requires consideration of nuclear quantum effects, resolving the interplay of these physical effects and how they are engendered in the diffusion mechanism has remained a subject of significant debate.

Early studies of proton transport in water invoked symmetry between the hydronium (H$_3$O$^+$) and hydroxide (OH$^-$) ions~\cite{huckel19283,eigen1964proton,eyring2012theoretical,zatsepina1972state,schioberg1973very,Librovich1979,khoshtariya1992new,Agmon2000} to suggest a framework where the structural transport mechanism of hydroxide ions could be inferred directly as the inverse of the corresponding mechanism for excess protons. However, it has since become clear that the two ions follow distinct proton transfer pathways, a phenomenon that is commonly attributed to differences in their solvation patterns~\cite{huckel19283,eigen1964proton,eyring2012theoretical,zatsepina1972state,schioberg1973very,Librovich1979,khoshtariya1992new,Agmon2000}. In particular, OH$^{-}$ can exhibit a hypercoordinated configuration where it accepts four hydrogen bonds~\cite{Tuckerman2002}, whereas the H$_3$O$^+$ ion can only donate three. This effect underscores the importance of water's complex hydrogen bond network in facilitating and ultimately influencing the rate of proton transport.

One of the most commonly invoked approaches to simulate the bond making and breaking that accompanies proton transfer in the structural diffusion of proton defects has been to perform computationally costly {\it ab initio} molecular dynamics (AIMD) simulations, where forces are obtained on the fly from electronic structure calculations. In addition, due to the light hydrogen nuclei involved, capturing a complete picture of the transport of proton defects requires including nuclear quantum effects (NQE) such as tunnelling and zero-point energy. {\it Ab initio} path-integral simulations include both of these effects and have been shown to be vital for the accurate description of the structure and transport of both H$_3$O$^+$ and OH$^-$~\cite{Tuckerman2002,Marx2006,Agmon2000,Napoli2018}. While imaginary-time path-integral simulations exactly include NQEs for structural properties, path-integral-based methods such as centroid molecular dynamics (CMD)~\cite{Cao1994,Jang1999} and ring polymer molecular dynamics (RPMD)~\cite{Craig2004,Habershon2013} have been shown to provide reliable approximate quantum dynamics for condensed phase systems. However, 30-100 replicas of a classical system are usually needed for path-integral simulations of aqueous systems at room temperature when using the most commonly employed second-order path integral discretization approach~\cite{Ceriotti2016}, increasing the computational cost by at least 30 times compared to AIMD simulations with classical nuclei. As such, {\it ab initio} path-integral molecular dynamics (AI-PIMD) simulations of the lengths required to sample many proton transfer events (on the order of nanoseconds) and hence reliably converge proton transport properties have traditionally been prohibitively costly. Recent path integral acceleration approaches~\cite{Markland2018} such as those that combine multiple time scale molecular dynamics~\cite{Tuckerman1992} and ring polymer contraction~\cite{Markland2008,Markland2008-2,Fanourgakis2009,Marsalek2016,Kapil2016} have made these timescales accessible for AI-PIMD simulations of 100's of picoseconds for systems of 300-500 atoms~\cite{Marsalek2017,Napoli2018}, albeit still at a formidable computational cost. 

The recent ability to perform condensed-phase simulations that combine electronic structure methods---most commonly density functional theory (DFT)---with path-integral methods has led to the identification of failures of the electronic structure treatments that were previously obfuscated when the nuclei were treated classically. Since the zero-point energy in the OH stretch provides additional energy equivalent to raising the temperature of that coordinate by $\sim$2000~K, performing AI-PIMD simulations of liquid water explores much higher-energy regions of the potential energy surface, such as long chemical bond extensions, which causes significant issues when lower-tier generalized gradient approximation (GGA) exchange correlation functionals are employed~\cite{Marsalek2016,Ceriotti2016,Gasparotto2016,Marsalek2017}. For example, when the nuclei are treated classically, spurious self-interaction in the revPBE-D3 GGA functional, which leads to an overly weak OH covalent bond, is fortuitously largely canceled out by the exclusion of NQEs, and hence the reintroduction of NQEs worsens the GGA functional's description of water~\cite{Marsalek2017}. While it has been shown that this deficiency can be alleviated by combining PIMD calculations with more costly hybrid functionals such as revPBE0-D3~\cite{Marsalek2017,Cheng2019}, the charged nature of proton defects is likely to exacerbate these issues further. Given the number of vital chemical processes which involve proton defects in nanoconfinement or at interfaces that typically require system sizes of more than 500 atoms and even longer timescales (multi-nanosecond) to average over the heterogeneity of the environment, performing converged AI-PIMD simulations of proton defects in these systems is likely to be impractical for the foreseeable future.

Machine-learned potentials (MLPs) have recently emerged as a compelling alternative to {\it ab initio} simulations~\cite{PhysRevLett.98.146401,Behler2011,Behler2014}. By training MLPs on the energies and/or forces obtained from {\it ab initio} calculations on a small number of suitably selected configurations (typically on the order of 1000s), MLPs have been shown to be able to interpolate, and in certain cases extrapolate, the {\it ab initio} potential energy surface over a wide range of conditions. While MLPs that can successfully model the reactive dynamics of protonated water clusters~\cite{KondatiNatarajan2015,Schran2019} and NaOH solutions~\cite{Hellstrm2016,Hellstrm2018} have previously been developed, they have not aimed to capture the behavior of both H$_3$O$^+$ and OH$^-$ concurrently. Here, we develop and introduce a training set sampled from {\it ab initio} simulations of excess protons, hydroxide ions, and proton-hydroxide recombination events and use it to train MLPs at the GGA (revPBE-D3) and hybrid (revPBE0-D3) levels of theory. We show that these MLPs can be used to simultaneously capture the properties of both types of proton defects in water, thus allowing the study of excess proton diffusion, hydroxide ion diffusion, water autoionization, and defect recombination processes. We utilize these MLPs to run both classical and path-integral AIMD simulations, allowing us to assess the role of different tiers of treatment of the electronic structure and NQEs in determining the mechanism of proton transport.

\section{Building the Machine-Learned Potential}

\subsection{Training Set Creation} \label{sec:training_set}
We utilized a training set of 37102 configurations, prepared by combining 4594 bulk water configurations randomly sampled from a previously reported dataset~\cite{Morawietz2018} with 32508 new configurations containing proton defects. The added proton defect configurations consist entirely of neutral frames of water molecules that contain a proton defect pair (both an excess proton and hydroxide ion). We do not include frames of water molecules containing the excess proton or hydroxide ion in isolation because such configurations require an opposite homogeneous background charge to neutralize the simulation box, which leads to box energies that vary depending on the box volume and Ewald summation parameters. The resulting variation in the energies complicates the fitting of an MLP, and hence we concentrate on fitting the MLP to the more physical neutral configurations where both the excess proton and hydroxide ion are present.

To prepare the training configurations that incorporate both excess protons and hydroxide ions, we selected frames from a revPBE-D3 AIMD trajectory of the hydroxide ion in water, identified the farthest water molecule from the OH$^-$, and added an excess proton to it, thereby neutralizing the frame. We used these frames to initialize classical and quantum revPBE-D3 AIMD trajectories with the aim of simulating proton defect recombination. From the resulting trajectories, we sampled configurations from the subset of frames where the proton defects had not recombined. Configurations were separately sampled to obtain a training set with near uniform distributions of the proton sharing coordinate ($\delta$, see section~\ref{subsec:traj_analysis} for definition) for the excess proton and hydroxide ion. Finally, similar to the configurations in the starting water dataset~\cite{Morawietz2018}, $\sim$67\% of the defect-separated frames were augmented by randomly displacing atoms to create configurations with higher forces, which served to improve model stability. The details of the training set are summarized in SI Table III.

Once the training and validation set configurations were obtained, their energies and forces were re-evaluated at the revPBE~\cite{Perdew1996,Zhang1998} and revPBE0~\cite{Adamo1999,Goerigk2011} levels of DFT with D3 dispersion~\cite{Grimme2010} using the CP2K program~\cite{VandeVondele2005,Khne2020}. Atomic cores were represented via the Godecker-Tetter-Hutter pseudopotentials~\cite{Goedecker1996}. We employed the hybrid Gaussian and plane wave density functional scheme~\cite{LIPPERT1997}, where the Kohn-Sham orbitals were expanded in the larger molecular optimized (MOLOPT)~\cite{VandeVondele2007} TZV2P basis set, and an auxiliary plane-wave basis was used to represent the density with a cutoff of 400 Ry for the revPBE-D3 calculation and 900 Ry for revPBE0-D3. Due to the relatively compact size of our training set, we are able to evaluate all of the configurations using a more accurate basis set that would be exceptionally computationally costly to use in AIMD simulations.

\subsection{Architecture and Training of the Machine-Learned Potential} \label{sec:mlp}
Our revPBE-D3 MLP was fit using the RuNNer package~\cite{RuNNer}, while the revPBE0-D3 MLP was fit using the n2p2 package~\cite{n2p2}. We employed Behler-Parrinello neural networks~\cite{PhysRevLett.98.146401,Morawietz2016} with two hidden layers containing 25 nodes each and an input layer containing 56 and 46 nodes for the H and O neural networks respectively. Chemical environments were described by radial and angular atom-centered symmetry function descriptors~\cite{Behler2011} with a cutoff of 6.35~$\text{\r{A}}$. We employed a 90/10 train/validation split, with 33425 configurations in the training set and 3677 configurations in the validation set. Training was done over 20 epochs, and MLP weights were fit to forces and energies. The final energy and force validation errors were 0.433 meV/atom and 65.8 meV/$\text{\r{A}}$ for the GGA MLP and 0.485 meV/atom and 39.4 meV/$\text{\r{A}}$ for the hybrid MLP respectively.

\section{Simulation Details}

\subsection{MD Simulations}
We performed classical and path integral simulations of both the excess proton and hydroxide ion in water under NVT conditions at T=300~K. The potential energy surfaces were described by MLPs trained on configurations from revPBE~\cite{Perdew1996,Zhang1998} (GGA) and revPBE0~\cite{Adamo1999,Goerigk2011} (hybrid) AIMD simulations with D3 dispersion~\cite{Grimme2010} (see subsections~\ref{sec:training_set} and~\ref{sec:mlp}), yielding four simulation protocols: classical GGA, classical hybrid, quantum GGA, and quantum hybrid. For all simulation protocols, we used a cubic box of length 15.66~$\text{\r{A}}$ with periodic boundary conditions. The simulation box contained 128 water molecules from which one proton was either removed, creating a hydroxide ion, or to which a proton was added, yielding an excess proton and resulting in a proton defect concentration in both cases of 0.43 M. To investigate finite-size effects, two sets of additional classical GGA trajectories were run in cubic boxes of lengths 19.73~$\text{\r{A}}$ and 24.86~$\text{\r{A}}$. In these one proton was added (excess proton) or removed (hydroxide ion) from simulations consisting of 256 and 512 water molecules, yielding proton defect concentrations of 0.22 M and 0.11 M respectively.

Classical MLP simulations were run with a 0.5~fs time step using the LAMMPS package~\cite{LAMMPS}, which employed n2p2~\cite{Singraber2019} to incorporate the MLP. A stochastic velocity rescaling (SVR) thermostat~\cite{Bussi2007} with a time constant of 1~ps was used to sample the canonical ensemble. For both the classical GGA and classical hybrid simulation protocols, we ran 107$\times$ and 70$\times$ 200-ps trajectories of the excess proton and hydroxide ion respectively at a box size of 15.66~$\text{\r{A}}$. This resulted in 21.4~ns of acid trajectory and 14~ns of base trajectory for each of the classical GGA and classical hybrid simulation protocols, with frames that were recorded every 2~fs. For the bigger cell sizes (boxes of lengths 19.73~$\text{\r{A}}$ and 24.86~$\text{\r{A}}$), we ran $\sim$100$\times$ and $\sim$80$\times$ 200-ps classical GGA trajectories of both the acid and base, resulting in 20~ns and 16~ns of trajectory respectively.

Path-integral MLP simulations were run with a 0.25~fs time step by employing the i-PI program~\cite{Ceriotti2014,Kapil2019}, which used the LAMMPS package~\cite{LAMMPS} (with n2p2~\cite{Singraber2019} used for the MLP) to compute energies and forces. The quantum path-integral simulations were performed via thermostatted ring polymer dynamics (TRPMD)~\cite{Craig2004,Habershon2013,Rossi2014} using 32 beads that were thermostatted with the path integral Langevin equation (PILE)~\cite{Ceriotti2010}. Under this scheme, ring polymer internal modes with frequency $\omega_k$ were subjected to a Langevin thermostat with friction $\gamma_k = 2 \lambda \omega_k$ and $\lambda=0.5$. For the quantum GGA and quantum hybrid simulation protocols, we ran 10 $\times$ 200-ps trajectories of both the excess proton and hydroxide ion, resulting in 2~ns of quantum trajectories for each combination of quantum simulation protocol and proton defect. Like in the classical case, frames were recorded every 2~fs.

We also performed classical GGA AIMD simulations of the excess proton and hydroxide ion in water under NVT conditions at T=300~K to serve as a benchmark for our MLP simulations. The AIMD simulations were run under periodic boundary conditions in cubic boxes of length 12.42~$\text{\r{A}}$ containing 64 water molecules from which a proton was removed or to which a proton was added. We employed the i-PI program~\cite{Ceriotti2014,Kapil2019} and its MTS~\cite{Tuckerman1992} implementation~\cite{Kapil2016} to propagate the nuclei. The full and reference forces were evaluated using the CP2K program~\cite{VandeVondele2005,Khne2020}. Full forces were computed at the revPBE~\cite{Perdew1996,Zhang1998} level of DFT with D3 dispersion~\cite{Grimme2010}. Atomic cores were represented via the Godecker-Tetter-Hutter pseudopotentials~\cite{Goedecker1996}. We employed the hybrid Gaussian and plane wave density functional scheme~\cite{LIPPERT1997}, where the Kohn-Sham orbitals were expanded in the TZV2P basis set, and an auxiliary plane-wave basis with a cutoff of 400 Ry was used to represent the density. The self-consistent field cycle was converged to an electronic gradient tolerance of $\epsilon_{\mathrm{SCF}} = 5 \times 10^{-7}$ using the orbital transformation method~\cite{VandeVondele2003} with the initial guess provided by the always stable predictor-corrector extrapolation method~\cite{Kolafa2003,Khne2007} at each AIMD step. Reference forces for the MTS were evaluated at the SCC-DFTB level in periodic boundary conditions using Ewald summation for electrostatics. The parametrizations for H and O atoms provided by CP2K were used and the D3 dispersion correction was added. The AIMD simulations were performed using an MTS propagator with the full force evaluated with a time step of 2.0~fs and the reference force with a time step of 0.5~fs. The SVR thermostat was employed with a time constant of 1~ps. We obtained total simulation times of 718~ps divided over 2 trajectories for the acid and 800~ps divided over 4 trajectories for the base.

\subsection{Trajectory Analysis} \label{subsec:traj_analysis}
A great deal of the results presented here arise from tracking the proton defect, which was identified at each frame by assigning every H atom to its nearest O and picking out the triply coordinated O atom (H$_3$O$^+$) in the excess proton trajectories and the singly coordinated O atom (OH$^-$) in the hydroxide ion trajectory. The atoms that make up the proton defect are referred to as O$^*$ and H$^*$ throughout this manuscript. Occasionally, highly transient water autolysis events (2H$_2$O $\rightarrow$ H$_3$O$^+$ + OH$^-$) would occur, forming excess ($>$1) proton defects at their respective frame. These events were rare, ranging from a maximum of 0.044\% of all frames for the quantum GGA excess proton trajectories to a minimum of 6.5$\times$ 10$^{-5}$\% of all frames for the classical hybrid excess proton trajectories. The ions resulting from these events were disregarded in our analysis, which instead focused on tracking the movement of the proton defect present at t=0. We defined the hydrogen bond geometrically as an atomic triplet O$_d$--H$_d$...O$_a$ (where O$_d$ and H$_d$ are covalently bonded donor atoms and O$_a$ is an acceptor atom) with $|$O$_d$O$_a$$|$ $\leq$ 3.5~$\text{\r{A}}$ and $\angle$ O$_a$O$_d$H$_d$ $\leq$ 30 $\degree$~\cite{Luzar1996}.

We computed mean square displacements (MSDs) for both the proton defect and water (O atoms) in our trajectories via the formula:
\begin{equation}
    \mathrm{MSD}(\Delta t) = \langle |\mathbf{r}(t_0 + \Delta t) - \mathbf{r}(t_0)|^2 \rangle,
\end{equation}
where $\langle \rangle$ is an ensemble average computed over all time origins $t_0$ and all relevant atoms. Diffusion coefficients were obtained by performing a linear fit to the MSD in the range 4~ps $\le \Delta t \le$ 20~ps and dividing it by $2d = 6$, where $d$ refers to the 3 dimensions of the simulation. We performed finite-size corrections for the water diffusion coefficient according to~\cite{Yeh2004}:
\begin{equation}
    D(\infty) = D(L) + \frac{\xi k_{\mathrm{B}}T}{6 \pi \eta L},
\end{equation}
where $L$ is the length of the simulation cell, $k_{\mathrm{B}}$ is the Boltzmann constant, and $T$ is the temperature, $\xi=2.837297$ is based on the cubic geometry of the simulation cell, and $\eta=0.8925 \times 10^{-3}$ Pa s is the experimental shear viscosity.

For the TRPMD simulations, we used the positions of the centroids to compute the MSD for the diffusion coefficient (which is a property of the long-time slope and hence gives the same result as using the beads); all other observables were calculated from the positions of the individual beads.

\section{Validation of the Machine-Learned Potential} \label{sec:validation}

We begin by evaluating how well classical molecular dynamics ran with our GGA-trained MLP reproduces observables from classical GGA {\it ab initio} molecular dynamics (AIMD) simulations. We benchmark on classical GGA since it has the lowest computational cost of the electronic structure and dynamics approaches presented in this study, and hence we can generate relatively long (718 ps for the acid and 800 ps for the base) AIMD trajectories with minimal statistical noise. Thus the discrepancies between the MLP and AIMD simulations in obtaining the properties of the systems discussed below, with the exception of the diffusion coefficients, do not arise from statistical sampling errors but rather either from errors in the MLP or due to the fact that the MLP was fit to a larger and more accurate MOLOPT basis set which was too computationally expensive to use for our long AIMD simulations. As shown in Fig.~\ref{fig:msd_dist}, even nanosecond-long trajectories lead to significant statistical error bars in the diffusion coefficients, so benchmarking the MLP on this property using AIMD is extremely challenging. This is one of the main motivations for the development of the MLP, which allows for the generation of trajectories that are long enough to converge this important property.

\begin{figure}[!ht]
    \centering
    \includegraphics[width=0.45\textwidth]{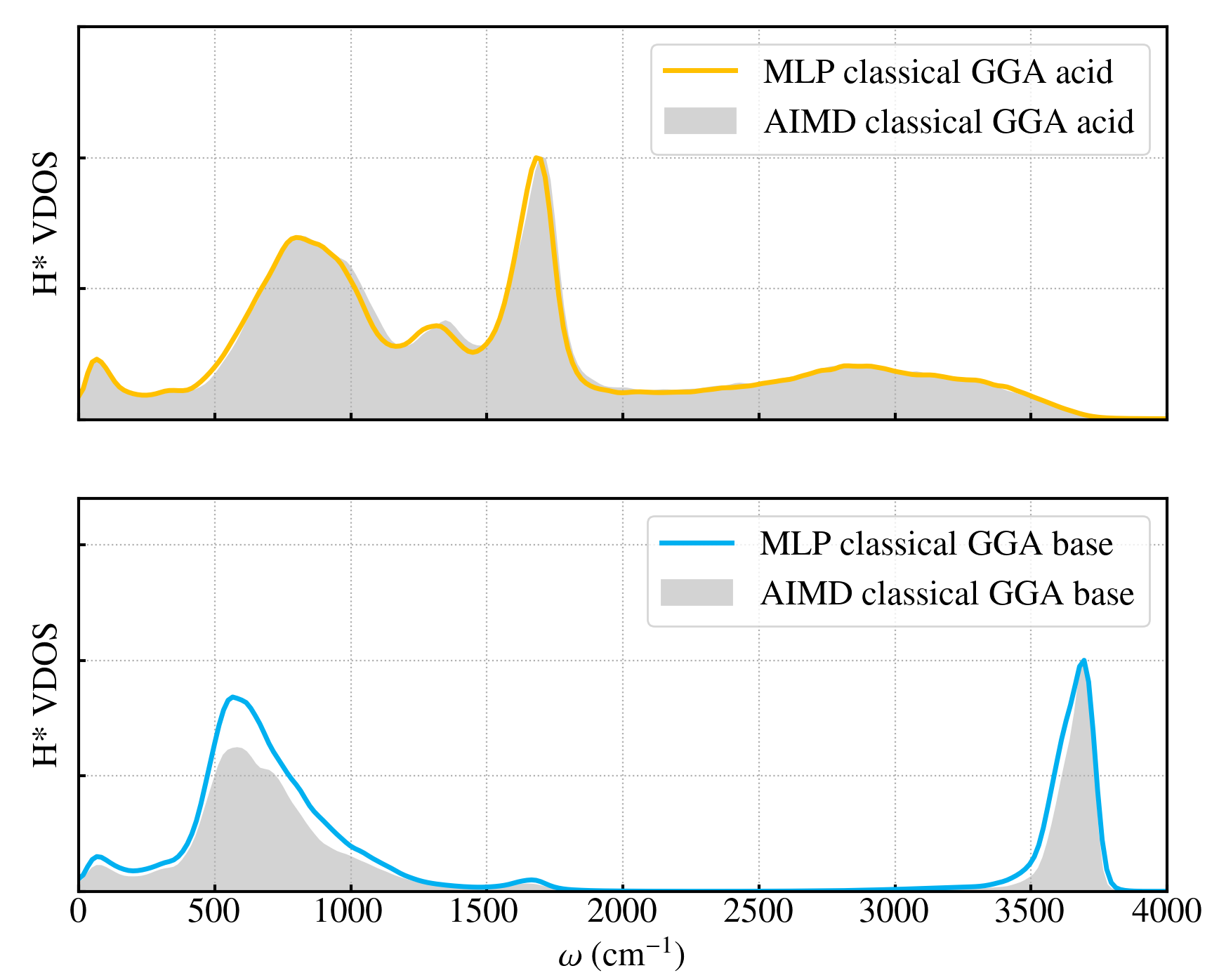}
    \caption{Comparison of the H$^*$ VDOS for revPBE-D3 AIMD and the revPBE-D3-trained MLP trajectories. The top panel shows this comparison for acid trajectories, while the bottom panel shows the comparison for base trajectories.}
    \label{fig:hs_vdos}
\end{figure}

\begin{figure}[!ht]
    \centering
    \includegraphics[width=0.45\textwidth]{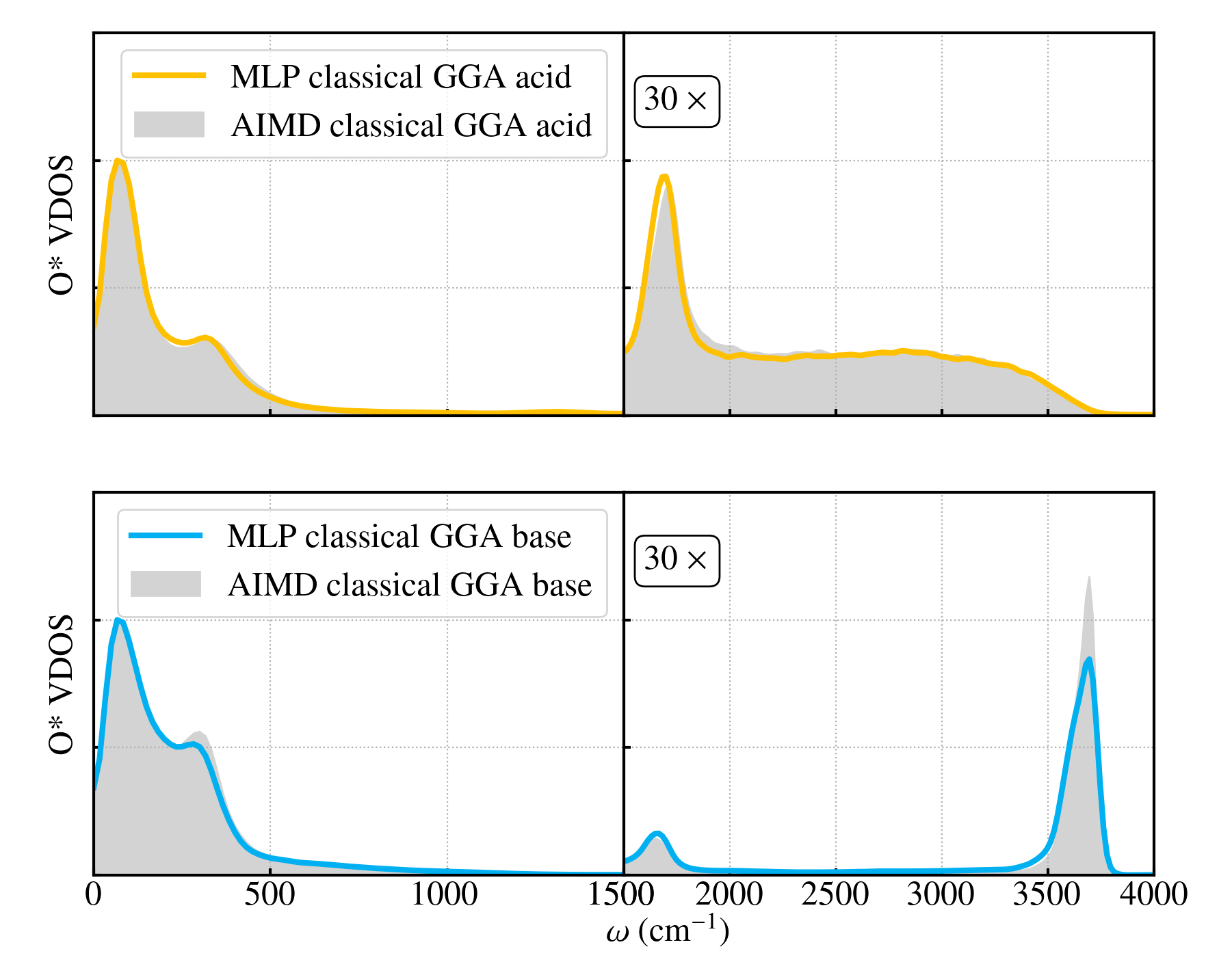}
    \caption{Comparison of the O$^*$ VDOS for revPBE-D3 AIMD and the revPBE-D3-trained MLP trajectories. The top panel shows this comparison for acid trajectories, while the bottom panel shows the comparison for base trajectories.}
    \label{fig:os_vdos}
\end{figure}

We consider the vibrational density of states (VDOS) for H$^*$ and O$^*$---i.e., the H and O atoms that make up the H$_3$O$^+$ and OH$^{-}$ defects in simulations with an excess proton and a hydroxide ion respectively (see Sec.~\ref{subsec:traj_analysis})---shown in Figs.~\ref{fig:hs_vdos} and~\ref{fig:os_vdos} respectively. These VDOS focus on the reactive defects and thus provide a stricter test of the MLP than the VDOS of all H and O atoms in the system (see SI Sec. II), most of which are bound to water molecules. For the excess proton, the MLP manages to accurately reproduce all features of the H$^*$ VDOS that are characteristic of the reactive defects~\cite{Fournier2018,Agmon2016}. These include the blue-shifted librational band at $\sim$800 cm$^{-1}$, the broadened bending mode peak at $\sim$1600 cm$^{-1}$, the broad absorption band between the bend and the OH stretch region (2000-3600 cm$^{-1}$), and a peak around $\sim$1250 cm$^{-1}$ which has previously been assigned to a Zundel-like proton shuttling motion~\cite{Headrick2004,Headrick2005,Asmis2003}. The H$^*$ VDOS for the hydroxide ion has fewer features: a librational band at $\sim$600 cm$^{-1}$, a comparatively smaller bending mode peak at $\sim$1600 cm$^{-1}$, and an OH stretch peak at $\sim$3600 cm$^{-1}$ that is sharper than that of pure liquid water~\cite{Roberts2009,Agmon2016}. While the MLP manages to accurately reproduce the frequencies of the features, it overestimates the amplitude of the librational band ($\sim$600 cm$^{-1}$) for the hydroxide ion. Figure~\ref{fig:os_vdos} shows quantitative agreement in the O$^*$ VDOS between the MLP and AIMD trajectories, with the much weaker bending mode peaks ($\sim$1600 cm$^{-1}$ for both the excess proton and hydroxide ion), the absorption band (2000-3600 cm$^{-1}$ for the excess proton), the stretch signal ($\sim$3600 cm$^{-1}$ for the hydroxide ion) and low-frequency features ($<$ 500 cm$^{-1}$) being faithfully reproduced by the MLP.

\begin{figure}[!ht]
    \centering
    \includegraphics[width=0.4\textwidth]{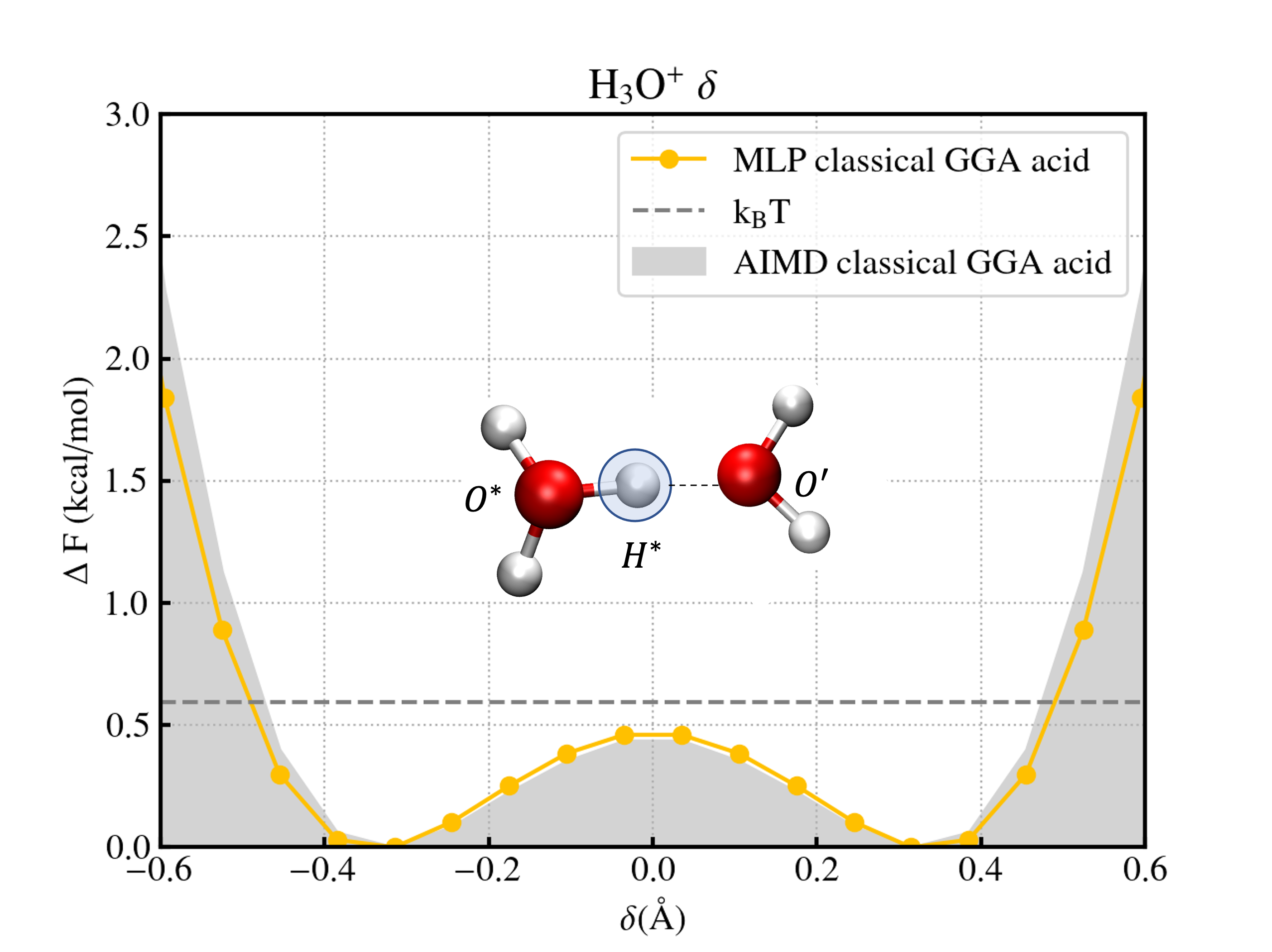}
    \caption{Comparison of the proton transfer free energy barrier along the proton sharing coordinate, $\delta = \mathrm{O^{\prime}H^* - O^*H^*}$, for revPBE-D3 AIMD and revPBE-D3-trained MLP acid trajectories. We only consider the H$^*$ with the lowest $\delta$ value at each frame.}
    \label{fig:acid_delta}
\end{figure}

\begin{figure}[!ht]
    \centering
    \includegraphics[width=0.4\textwidth]{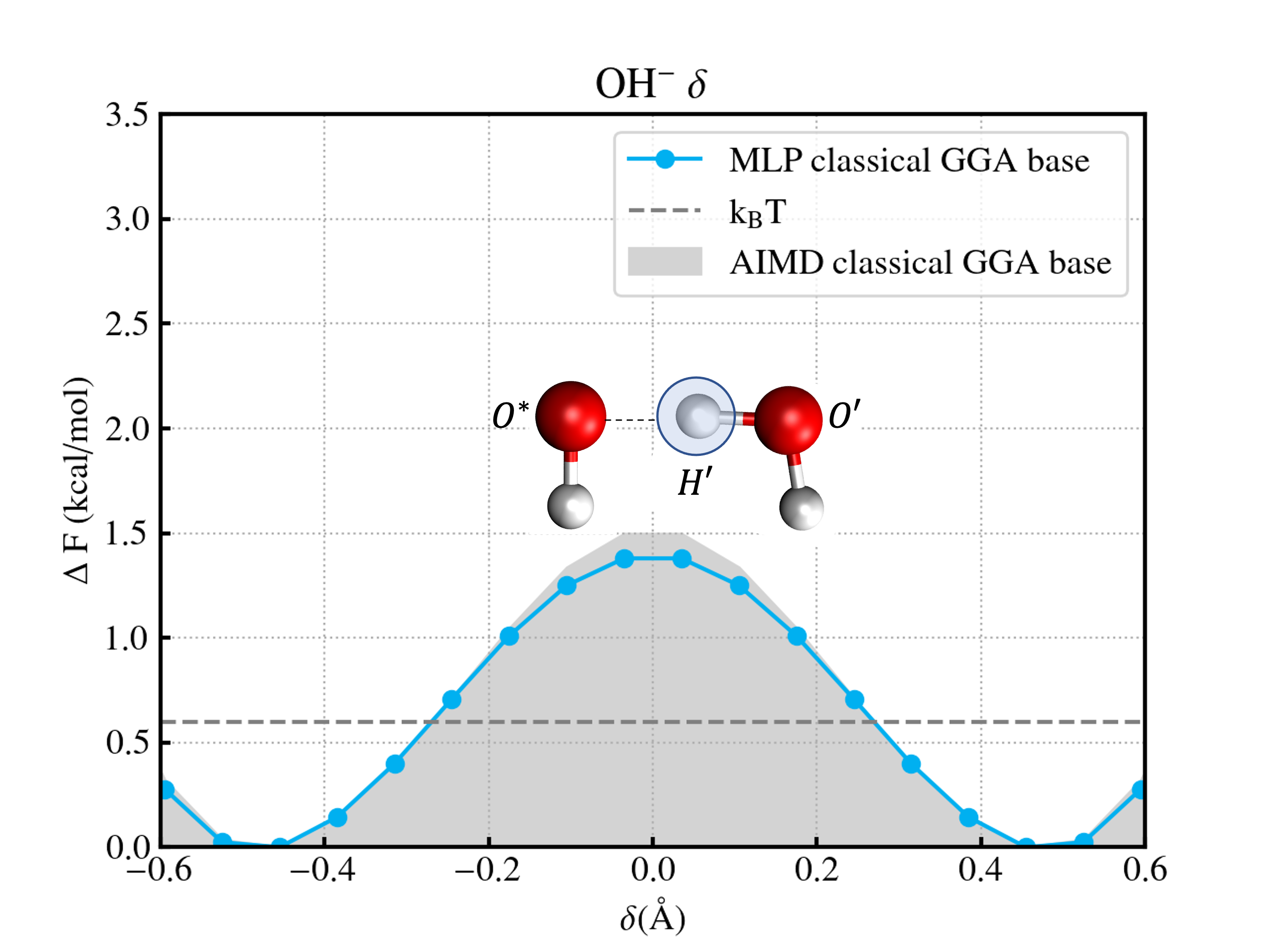}
    \caption{Comparison of the proton transfer free energy barrier along the proton sharing coordinate, $\delta = \mathrm{O^{*}H^{\prime} - O^{\prime}H^{\prime}}$, for revPBE-D3 AIMD and revPBE-D3-trained MLP base trajectories. We only consider the H$^{\prime}$ closest to the O$^*$ at each frame.}
    \label{fig:base_delta}
\end{figure}

In Figures~\ref{fig:acid_delta} and~\ref{fig:base_delta}, we evaluate how well the GGA-trained MLP reproduces the AIMD free energy along the proton sharing coordinate. As illustrated in Figs.~\ref{fig:acid_delta} and~\ref{fig:base_delta}, $\delta = d_{\mathrm{O^{\prime}H^*}} - d_{\mathrm{O^*H^*}}$ for the excess proton and $\delta = d_{\mathrm{O^{*}H^{\prime}}} - d_{\mathrm{{O^{\prime}H^{\prime}}}}$ for the hydroxide ion, where $d$ denotes the distance between the respective atoms. O$^*$ and H$^*$ are the defect atoms defined above, while O$^{\prime}$ and H$^{\prime}$ are atoms in the first solvation shell of the charge defect. For the excess proton simulations, of the $\delta$ values from the three H$^*$ atoms connected to O$^*$, only the lowest was used, and for the hydroxide ion simulation, the $\delta$ values were calculated based only on the H$^{\prime}$ closest to O$^*$. The free energy along the delta coordinate $\Delta F(\delta) = -k_{\mathrm{B}}T \ln P(\delta)$ was calculated from the resulting $\delta$ probability distribution, $P(\delta)$. The two free energy minima along the proton sharing coordinate, $\delta$, thus correspond to the covalent bonding of the hydrogen atom to one or the other oxygen atom, and the height of the free energy barrier between the two minima is located at $\delta$=0 due to the symmetry of the coordinate.

Figures~\ref{fig:acid_delta} and~\ref{fig:base_delta} show that the MLP simulations accurately reproduce both the positions of the free energy minima and height of the free energy barrier at $\delta=0$, $\delta F(\delta=0)$, obtained from AIMD, with the MLP overestimating it by 0.02 kcal/mol for the acid and underestimating it by 0.12 kcal/mol for the base. At 300~K, these errors correspond to 0.03 and 0.2 $k_{\mathrm{B}}T$ respectively, are much smaller than the thermal energy in the system, and are completely dwarfed by the zero-point energy along these coordinates as discussed further in Sec.~\ref{sec:results}. 

To further validate that the structural and dynamical properties of the excess proton and hydroxide ion in liquid water are captured by our GGA MLP, we show in SI Sec. I that the MLP also quantitatively reproduces the AIMD O$^*$-O, O$^*$-H, and H$^*$-H radial distribution functions for both simulations. Finally, the MLP trajectories yield O$^*$ diffusion coefficients of 8.04 $\times$ 10$^{-9}$ and 4.95 $\times$ 10$^{-9}$ m$^2$/s for the excess proton and hydroxide ion respectively, which compare favorably to corresponding AIMD values of 9.87 $\times$ 10$^{-9}$ and 3.13 $\times$ 10$^{-9}$ m$^2$/s. As discussed in Sec.~\ref{sec:results}, the slow convergence of the proton defect diffusion coefficient with simulation time is such that the difference between the MLP and AIMD diffusion coefficients can be accounted for by statistical uncertainty. This is further illustrated in SI Fig. 7.

\section{Results} \label{sec:results}

\begin{figure}[!ht]
    \centering
    \includegraphics[width=0.4\textwidth]{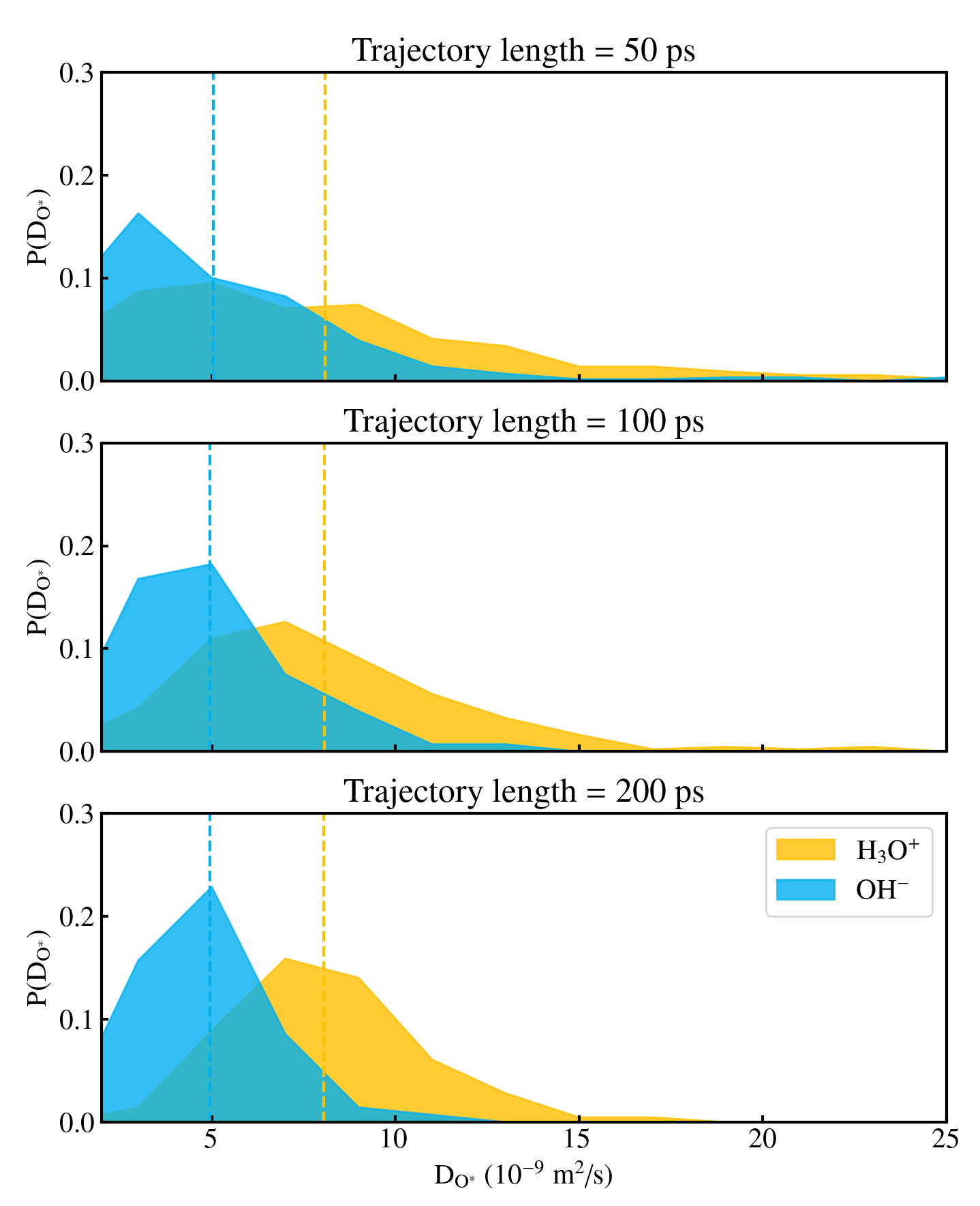}
    \caption{Proton defect diffusion coefficient distributions (P(D$_{\mathrm{O^*}}$)) of the MLP GGA simulations computed at different trajectory lengths, with the corresponding means displayed as dashed lines. The degree of separation between the D$_{\mathrm{O^*}}$ distributions for the acid and base increases with the trajectory length, underscoring the importance of computing D$_{\mathrm{O^*}}$ values from trajectories that are at least hundreds of picoseconds long.}
    \label{fig:msd_dist}
\end{figure}

Having established the accuracy of the GGA MLP trained on {\it ab initio} data in the previous section, we now investigate the properties of excess protons and hydroxide ions in water at the GGA and hybrid levels of DFT with and without nuclear quantum effects. Due to the low computational cost of evaluating MLPs, we investigate properties (such as the proton defect diffusion coefficients) that require multiple nanoseconds of simulation to reliably converge. The importance of using such long simulations is illustrated in Fig.~\ref{fig:msd_dist}, which shows the distribution of diffusion coefficients obtained from 50~ps, 100~ps, and 200~ps trajectory segments derived from subdividing our total of 20~ns of revPBE-D3 (GGA) MLP trajectory. 
A single simulation performed on a 50~ps, 100~ps, or 200~ps timescale would thus correspond to picking a single realization from these distributions (with each diffusion coefficient corresponding to the linear fit to one of the MSD curves shown in SI Figures 9 and 10, as detailed in Section~\ref{subsec:traj_analysis}). As one can see, even with a 200~ps simulation, a time longer than that in many previous AIMD studies of proton defects, the wide distribution of diffusion coefficients that can be obtained does not allow one to reliably distinguish between the higher expected diffusion coefficient of an excess proton and the lower one expected for the hydroxide ion. This emphasizes the need for multiple-nanosecond simulations to comment on the relative transport rates of the excess proton and hydroxide ion in liquid water.

\begin{figure*}[!ht]
    \centering
    \includegraphics[width=0.7\textwidth]{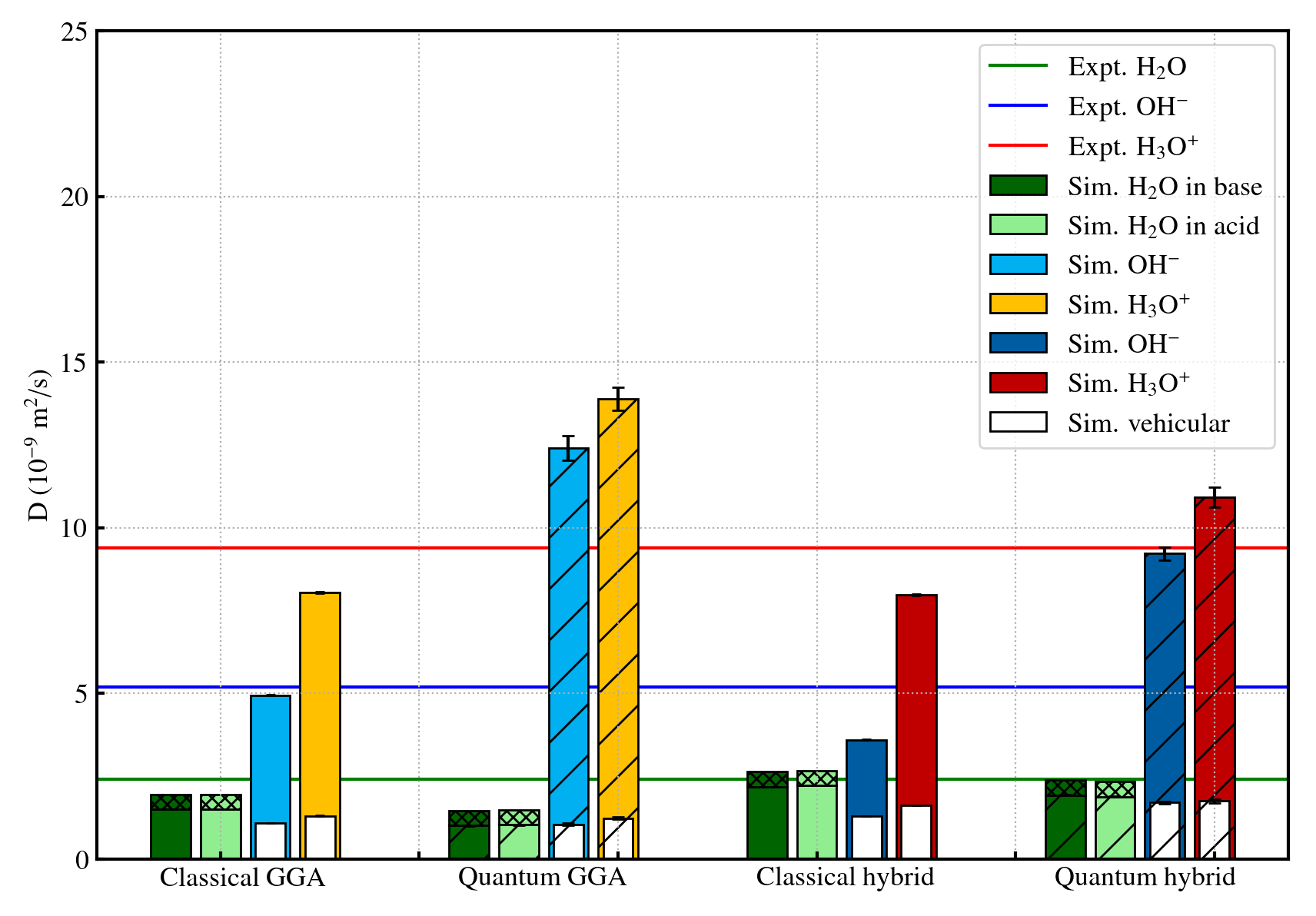}
    \caption{Acid and base diffusion coefficients calculated from the MLP trajectories. We report six values for each MLP run: the molecular diffusion coefficient of water (measured by tracking O atoms) in the acid and base, the proton defect diffusion coefficient (measured by tracking O$^*$) in the acid and base, and the vehicular component of the proton defect diffusion coefficient for the acid and base. Experimental values for the diffusion coefficients of water~\cite{Holz2000}, H$_3$O$^+$~\cite{Sluyters2010}, and OH$^-$~\cite{Sluyters2010} are shown for comparison.}
    \label{fig:d_vals}
\end{figure*}

Figure~\ref{fig:d_vals} summarizes the water molecule and proton defect diffusion coefficients calculated from classical and quantum (TRPMD) simulations of MLPs trained on revPBE-D3 (GGA) and revPBE0-D3 (hybrid) AIMD simulations. For each combination of dynamics approach (classical or quantum) and electronic structure approach (GGA or hybrid), we performed separate simulations of an excess proton in water and a hydroxide ion in water, i.e., simulations of a water box that initially contained 128 water molecules to which one proton has been added or from which one proton has been removed. Each set of four bars shown in Fig.~\ref{fig:d_vals} corresponds to a particular simulation protocol, i.e., a combination of dynamics and electronic structure approaches, and within each set, the colored bars correspond to the diffusion coefficients of water molecules in the basic solution (dark green), water molecules in the acidic solution (light green), and the proton defect diffusion coefficients for the hydroxide ion (blue/navy) and the excess proton (yellow/red) respectively. 

From Fig.~\ref{fig:d_vals}, one can see that in all cases, proton defect diffusion coefficients are significantly higher than those of water molecules and that the excess proton diffuses faster than the hydroxide ion, which is in line with the experimentally observed trend~\cite{Sluyters2010}. The solid horizontal lines in Fig.~\ref{fig:d_vals} show the experimentally observed diffusion coefficients of the relevant species~\cite{Holz2000,Sluyters2010}. Similar to previous studies, the water molecule diffusion coefficients obtained from revPBE-D3 and revPBE0-D3 AIMD simulations of pure water are in good agreement with the experimentally observed value when finite-size corrections~\cite{Yeh2004} are applied~\cite{Marsalek2017}. In Fig.~\ref{fig:d_vals}, finite-size corrections are shown as the hatched regions of the green bars and were computed using the experimental shear viscosity. For both the excess proton and the hydroxide ion simulations, the diffusion coefficients of the water molecules are within 0.1 $\times$ 10$^{-9}$ m$^2$/s of each other, indicating that at this low proton defect concentration (0.43 M), the defect has only a minor effect on the diffusion of the molecules in the liquid. 

For the water molecules, the hybrid functional exhibits faster diffusion than the GGA, and within a given choice of functional, the quantum simulations show slightly slower diffusion than the classical ones. The former observation can be rationalized as due to the hybrid functional's partial taming of the delocalization error inherent in (GGA) DFT, which alleviates the stronger hydrogen bonds~\cite{Gillan2016} and slower diffusion observed under GGA. The latter observation of slower diffusion upon including NQEs arises from the subtle balance of competing quantum effects in liquid water~\cite{Habershon2009,Chen2003,Markland2012} and other hydrogen-bonded systems~\cite{Li2011,McKenzie2012,McKenzie2014}, which in the case of DFT water~\cite{Guidon2008,Zhang2011,DiStasio2014,Miceli2015} and the revPBE-D3 and revPBE0-D3 functionals~\cite{Marsalek2017} has generally led to a slight structuring of the liquid and corresponding lowering of the diffusion coefficient. We note that due to the subtle cancellation of NQEs in liquid water at 300~K, NNPs~\cite{Chen2023} and other potentials~\cite{Babin2013,Babin2014,Medders2014,Reddy2016,Yu2022} fit to higher-level electronic structure methods such as CCSD(T) and AFQMC have shown a slight increase in water's diffusion coefficient upon treating the nuclei quantum mechanically. Of all the MLPs, the quantum hybrid trajectory most closely reproduces the experimental water diffusion coefficient, with computed system size-corrected diffusion coefficients of 2.33 $\times$ 10$^{-9}$ m$^2$/s and 2.38 $\times$ 10$^{-9}$ m$^2$/s for the acid and base respectively, compared to the experimental value of 2.41 $\times$ 10$^{-9}$ m$^2$/s~\cite{Holz2000}.

The diffusion coefficients of the excess proton (yellow/red) and hydroxide ion (blue/navy) are shown in Fig.~\ref{fig:d_vals}, with the horizontal red and blue lines showing the experimental diffusion coefficients. For both acid and base trajectories, the charge defect diffusion coefficient follows the trend: quantum GGA $>$ quantum hybrid $>$ classical GGA $>$ classical hybrid. These trends for the defect diffusion are roughly the opposite of what is seen for water molecules, with the GGA giving rise to faster defect diffusion than the hybrid and nuclear quantum effects also increasing the diffusion rate. The white bars in Fig.~\ref{fig:d_vals} show the vehicular component of the diffusion obtained by decomposing the total diffusion coefficients into their structural components, which arise entirely from intermolecular proton transfer events, and their vehicular components, which arise from the molecular motion of the proton defects (see SI Sec. III). We observe that in all cases, the vehicular component is a small and nearly constant part of proton defect diffusion and hence the changes in the total diffusion coefficients arise from variations in the dominant structural diffusion mechanism upon changing the exchange-correlation functional or including NQEs.

\begin{figure}[!ht]
    \centering
    \includegraphics[width=0.35\textwidth]{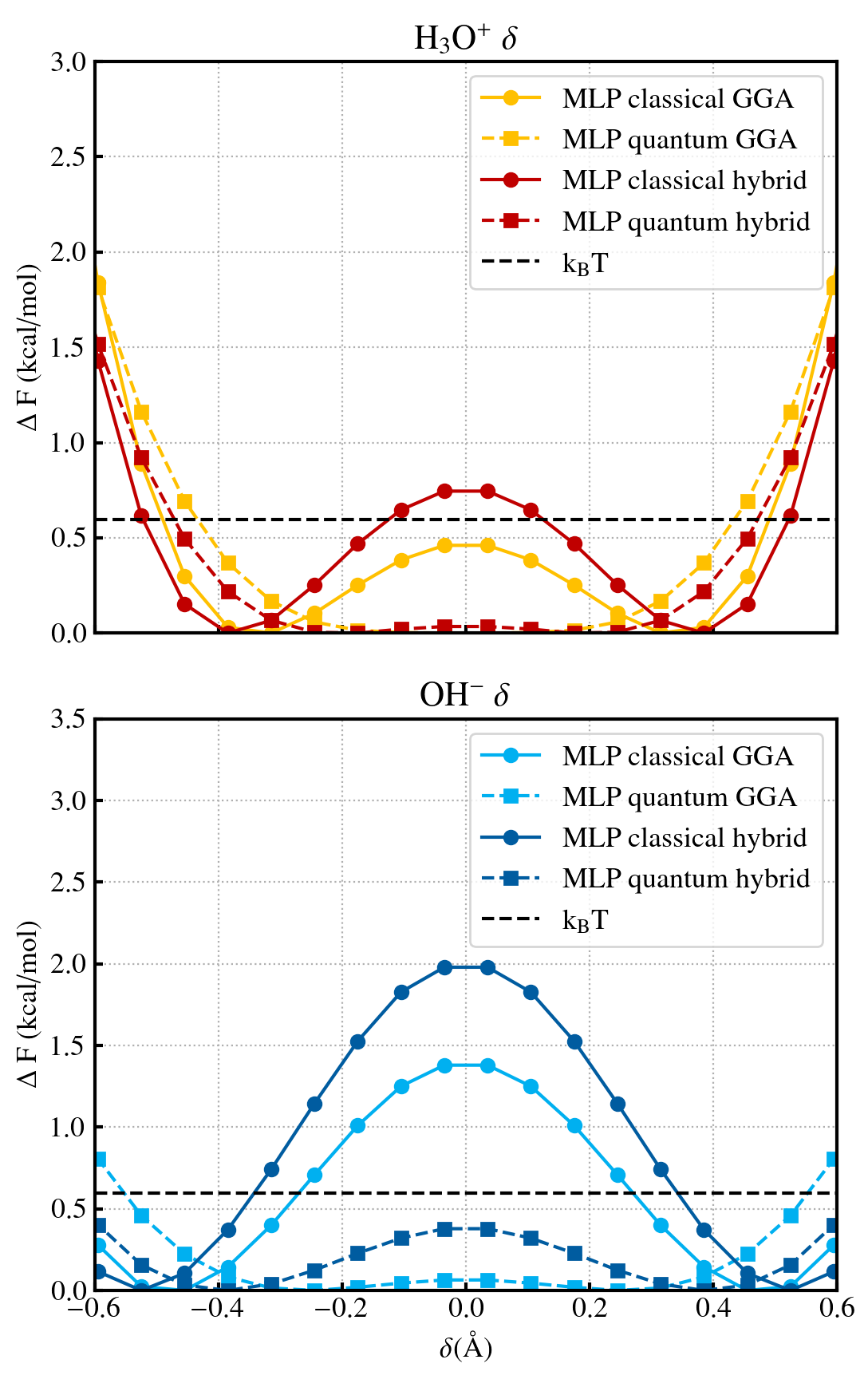}
    \caption{Free energy, $\Delta$F, along the proton sharing coordinate $\delta$, for the acid (top) and base (bottom) obtained from MLP simulations at 300~K. The dashed line shows $k_{\mathrm{B}}T$ at 300~K.}
    \label{fig:all_delta_F}
\end{figure}

To understand the origins of the trends in the rate of diffusion of proton defects in water, we begin by analyzing the free energy barrier for proton transfer under the different simulation protocols. Fig.~\ref{fig:all_delta_F} shows the free energy profile along the proton sharing coordinate, $\Delta F(\delta)$, defined for the excess proton and hydroxide ion in Sec.~\ref{sec:validation}. The height of the free energy barrier, $\Delta F(\delta=0)$, is larger for the hydroxide ion than for the excess proton, which is consistent with the slower diffusion of hydroxide ions compared to excess protons. For classical nuclei, which exhibit the most pronounced barrier, the difference in the barrier height between the two types of proton defect is 0.92 kcal/mol for the GGA and 1.23 kcal/mol for the hybrid functional. The proton transfer barrier obtained from the hybrid functional is higher than that of the GGA functional for both the excess proton and hydroxide ion by 0.28 kcal/mol and 0.60 kcal/mol respectively when a classical description of the nuclei is used. This behavior follows from the fact that the top of the barrier at $\delta$=0 corresponds to a scenario where the proton is equidistant from two O atoms and is thus a state of large charge separation. Due to the delocalization error in DFT charge, separated states under GGA are spuriously lowered in energy relative to charge-localized states, which decreases the free energy barrier. The hybrid functional somewhat alleviates this issue and, in turn, raises the free energy barrier along $\delta$. For both the excess proton and hydroxide ion, the minima in $\Delta F(\delta)$ obtained from the classical simulations are shifted closer to $\delta=0$ for GGA than the hybrid, indicating a more shared equilibrium position of the proton for the former.

Nuclear quantum effects are expected to play a major role in determining the free energy profile along the proton-sharing coordinate, which describes the movement of a light (hydrogen) atom across an energy barrier. The zero-point energy of an O-H stretch in water ($\hbar\omega$, $\omega$=3600 cm$^{-1}$) is 5.15 kcal/mol, which for the excess proton, is larger than the free energy barriers of 0.46 kcal/mol and 0.75 kcal/mol along the proton sharing coordinate obtained from both the classical GGA and hybrid simulations respectively. Upon including NQEs, the free energy barrier for the excess proton is whittled down to 0~kcal/mol for the GGA simulation and reduced to 0.03 kcal/mol for the hybrid simulation. In the case of the hydroxide ion, the classical free energy barriers of 1.38 kcal/mol and 1.98 kcal/mol for the GGA and hybrid simulations are substantially reduced by 1.32 kcal/mol and 1.60 kcal/mol respectively upon including NQEs. We note that these reductions in the free energy barrier are considerably smaller than the reduction that might be estimated by considering the ZPE along that coordinate, emphasizing that the mechanism of proton transport in solution is not fully captured by motion along this single coordinate.

\begin{figure}[!ht]
    \centering
    \includegraphics[width=0.35\textwidth]{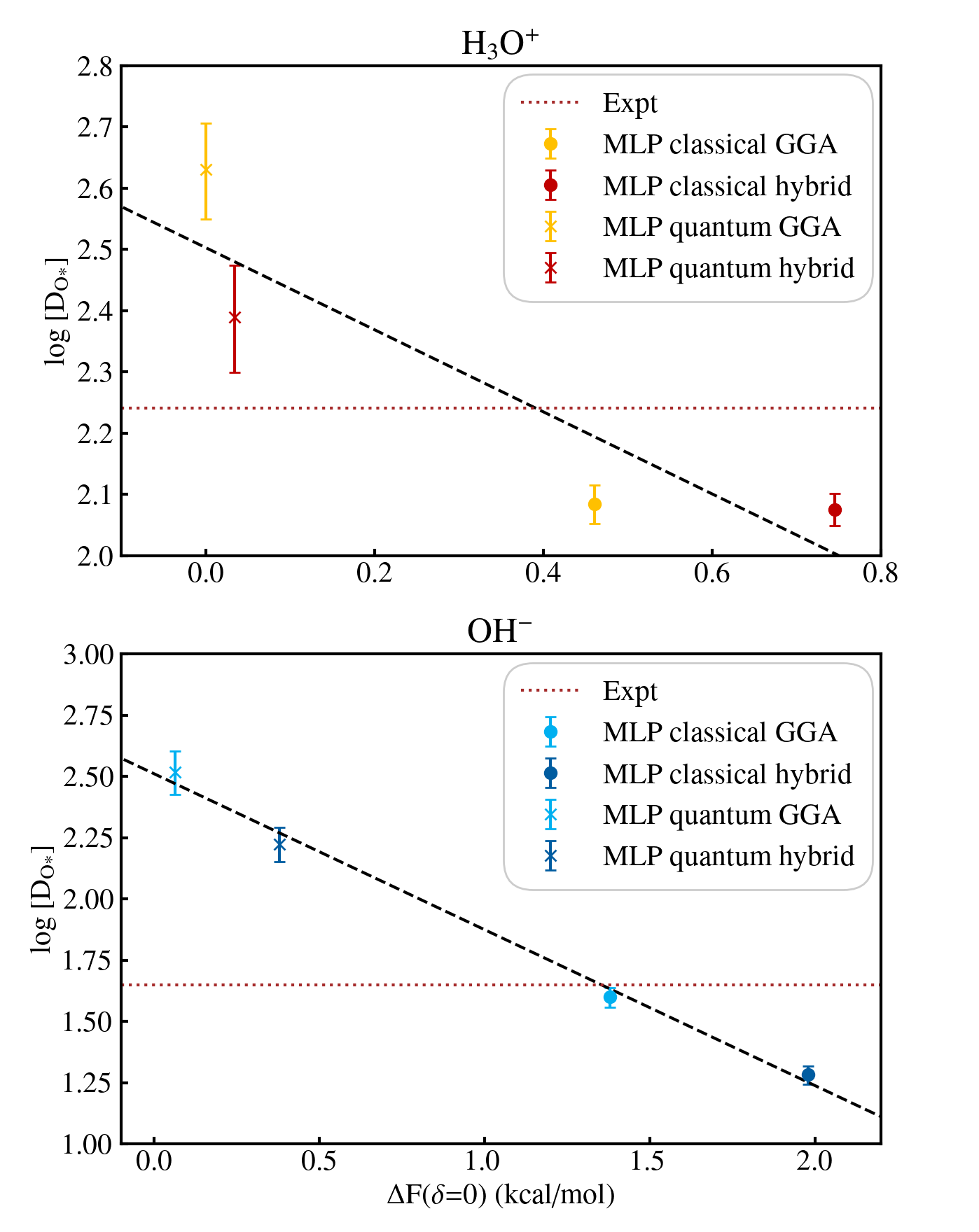}
    \caption{Diffusion coefficients as a function of the proton sharing coordinate free energy barrier, $\Delta$ F $(\delta=0)$, for the excess proton (top) and hydroxide ion (bottom) obtained from our MLP simulations. Our data show a much stronger correlation for the base than for the acid.}
    \label{fig:coeff_vs_delta}
\end{figure}

It is instructive to investigate how the variation observed in the height of the free energy barrier along the proton sharing coordinate ($\Delta F(\delta=0)$) under different simulation protocols correlates with the proton defect diffusion coefficients, $\mathrm{D_{O^{*}}}$. Due to the commonly observed exponential dependence of rates of processes on their associated free energy barrier, Fig.~\ref{fig:coeff_vs_delta} plots the natural logarithm of the defect diffusion coefficients $\log[\mathrm{D_{O^{*}}}$] against the free energy barrier along the proton sharing coordinate $\Delta F(\delta=0)$, which should give a linear relationship. As expected, there is an inverse linear relationship between them, i.e., an increase in the free energy barrier along the proton sharing coordinate decreases the likelihood of intermolecular proton hops and thus inhibits the transport of the proton defect. This inverse relationship is much stronger for the base than for the acid, which suggests that supramolecular factors beyond the intermolecular proton transfer barrier play a bigger role in the transport of H$_3$O$^+$.

Comparing the simulated diffusion coefficients to their experimental values obtained from conductivity data at 301 K for H$_3$O$^+$ (9.4 $\times$ 10$^{-9}$ m$^2$/s) and OH$^-$ (5.2 $\times$ 10$^{-9}$ m$^2$/s)~\cite{Sluyters2010}, both the classical GGA and quantum hybrid H$_3$O$^+$ simulations (8.0 $\times$ 10$^{-9}$ and 10.9 $\times$ 10$^{-9}$ m$^2$/s) most closely reproduce the experimental H$_3$O$^+$ diffusion coefficient, while the classical GGA OH$^-$ simulations (4.9 $\times$ 10$^{-9}$ m$^2$/s) most closely reproduce the experimental OH$^-$ diffusion coefficient. The strong performance of the classical GGA trajectories can likely be attributed to the cancellation of error between proton delocalization due to overestimated hydrogen bond strengths and proton localization due to the classical treatment of nuclei. Quantum hybrid trajectories also perform relatively well because they incorporate NQEs, and the revPBE0-D3 functional less severely overestimates hydrogen bond strengths. Notably, when NQEs are included for both the GGA and hybrid functionals, the ratios of the excess proton diffusion coefficient to that of the hydroxide ion diffusion coefficient, 1.1 and 1.2 respectively, are lower than the experimentally observed value of 1.8, and are in worse agreement for this property than when the nuclei are treated classically (1.6 and 2.2 for the GGA and hybrid simulations respectively). This arises from the much more pronounced NQEs on the hydroxide diffusion coefficient than the excess proton, with a factor of 2.5 and 2.6 increase for the GGA and hybrid upon going from classical to quantum for the hydroxide ion but only 1.7 and 1.4 for the excess proton.

\begin{figure*}[!ht]
    \centering
    \includegraphics[width=0.95\textwidth]{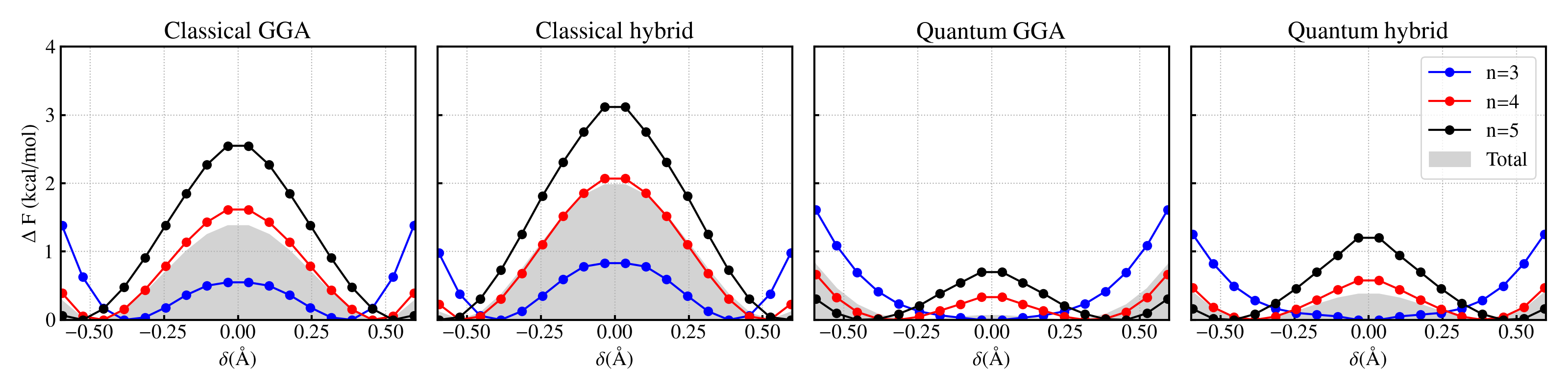}
    \caption{Values for the free energy barrier along the proton sharing coordinate $\delta$ at 3, 4, and 5 accepted hydrogen bonds at the OH$^-$ for all MLP trajectories. The $\Delta$F($\delta$) for all OH$^-$ is also shown for each trajectory.}
    \label{fig:delta_coord}
\end{figure*}

To further analyze trends in the hydroxide ion diffusion coefficient and the impact of NQEs, we now explore the relationship between the rate of diffusion of OH$^-$ and the proton transfer free energy barrier, $\Delta F(\delta=0)$. Previous studies have suggested that OH$^-$ predominantly exists in an inert hypercoordinated state where it accepts four hydrogen bonds and transiently donates one~\cite{Tuckerman2006,Chen2018}. In this picture, proton transfer occurs after thermal fluctuations break one of the accepted hydrogen bonds, converting the inert hypercoordinated OH$^-$ to a tetrahedral active state that accepts three hydrogen bonds and donates one. The OH$^-$ is then ready to accept a proton from a neighboring molecule because it has assumed the tetrahedral geometry typical of neutral water molecules, a concept known as presolvation~\cite{Tuckerman2006}. The hypercoordination of OH$^-$ is the primary reason why the mechanism of transport of OH$^-$ cannot simply be inferred from that of H$_3$O$^+$, which typically donates three hydrogen bonds. 

Our MLP simulations support the hypercoordination picture, with the less active state where OH$^-$ accepts four hydrogen bonds making up the majority of frames for all of the trajectories, i.e. 63\%, 56\%, 50\%, and 58\% of all frames in the classical GGA, classical hybrid, quantum GGA, and quantum hybrid trajectories respectively. In all cases, the percentage of all proton hops where the recipient is a triply-coordinated OH$^-$ is higher than the percentage of all OH$^-$ configurations that are triply coordinated. In particular, for the classical simulations, 51\% and 44\% (GGA and hybrid) of all proton hops are to triply-coordinated OH$^-$ while only 13\% and 7\% of all OH$^-$ configurations are triply coordinated. A similar trend is observed for the quantum simulations, where 62.6\% and 51.5\% (GGA and hybrid) of all proton hops are to triply-coordinated OH$^-$, while only 41\% and 20\% of all OH$^-$ configurations are triply coordinated. Further analysis shows that the hydroxide ion diffusion coefficient is positively correlated with the percentage of proton hops to a triply-coordinated OH$^-$. Additionally, there is a clear inverse correlation between the free energy barrier along the proton sharing coordinate and the OH$^-$ coordination number. Figure~\ref{fig:delta_coord} shows the computed free energy barriers for OH$^-$ at different numbers of accepted hydrogen bonds (n=\{3,4,5\}). OH$^-$ ions that accept three hydrogen bonds have the lowest $\Delta$F across all trajectories, further suggesting that the n=3 state is indeed the active proton transfer state, in line with previous studies\cite{Tuckerman2006,Chen2018}. 

We observe that the impact of NQEs on the proton transfer barrier is two-fold: the direct effect of lowering the barrier along the proton sharing coordinate and the indirect effect of shifting the distribution of hypercoordinated states towards the more favorable states for proton transfer to occur. Specifically, in Fig.~\ref{fig:delta_coord} one sees that $\Delta F(\delta=0)$ is lower for the quantum trajectories for any given value of the OH$^-$ coordination (e.g. n=3, n=4, etc.). This direct effect lowers the barrier by as much as 1.92 kcal/mol in the n=5 state and by 0.83 kcal/mol in the n=3 state. Conversely, NQEs provide an indirect effect by markedly increasing the incidence of triply coordinated OH$^-$ configurations: 12.6\%, and 6.7\% of all frames for classical GGA and hybrid trajectories respectively, compared to 41\% and 20\% of all frames in the quantum GGA and hybrid trajectories respectively.

\section{Conclusion}
We have presented two MLPs---trained on revPBE-D3 (GGA) and revPBE0-D3 (hybrid) AIMD energies and forces---that simultaneously capture the properties of excess protons and hydroxide ions in water. To test the validity of our MLPs, we benchmarked the GGA MLP against independent GGA AIMD trajectories of an excess proton in water and a hydroxide ion in water. Overall, the GGA MLP faithfully reproduced several of the most challenging {\it ab initio} properties relevant to proton defects, namely the H$^*$ and O$^*$ VDOS, the free energy barrier along $\delta$, and the RDFs (H$^*$--H, O$^*$--H, O$^*$--O).

Our hybrid and GGA MLPs were then used to perform multi-nanosecond classical and TRPMD trajectories of the excess proton and hydroxide ion, enabling us to obtain diffusion properties of the proton defects with minimal statistical noise. By analyzing these simulations, we elucidated how the choice of DFT functional (GGA vs hybrid) and nuclear representation (classical vs quantum) affects the rate of both molecular and proton defect diffusion. By comparing the proton defect diffusion coefficient to the free energy barrier along the proton sharing coordinate ($\delta$), we showed that a higher free energy barrier is correlated with a low rate of proton transfer, although the correlation is stronger for the hydroxide ion than for the excess proton. Additionally, by calculating the free energy barrier along $\delta$ for different coordination states of the OH$^-$ ion, we showed that our data agree with prior studies~\cite{Tuckerman2006,Chen2018} that posit a predominantly inert quadruply hydrogen-bonded OH$^-$ that occasionally undergoes thermal fluctuations to lose one of its accepted hydrogen bonds in order to enter a tetrahedral state that is conducive to proton transfer. 

The MLP models we introduce here provide a means to run multi-nanosecond molecular dynamics simulations of proton defects in water at DFT-level accuracy and low computational cost, thus enabling one to study rare events with unprecedented statistical accuracy. In addition, the training set constructed in this work provides a starting point for training MLPs that are able to treat both proton and hydroxide defects, and hence processes such as autoionization, at higher levels of electronic structure theory or for more diverse chemical environments. This lays the groundwork for improving our understanding of the finer details of the proton transfer mechanism in water, as well as the mechanics of autoionization and proton-hydroxide recombination events.

\section*{Acknowledgments}
This work was supported by National Science Foundation Grant No. CHE-2154291 (to T.E.M.). A.O.A. acknowledges support from the Stanford Diversifying Academia, Recruiting Excellence Fellowship. O.M. acknowledges support from the Czech Science Foundation, project No. 21-27987S. T.M. is grateful for financial support by the DFG (MO 3177/1-1).

\section*{Data Availability}
The data that supports the findings of this study are available within the article and its supplementary material.

\section{References}
\bibliography{references}
	
\end{document}